\documentclass[prl,twocolumn,epsf,psfig]{revtex4}
\usepackage{graphicx}
\DeclareGraphicsExtensions{.pdf,.png,.gif,.jpg}
\usepackage{float}
\usepackage{subfig}
\usepackage{color}
\usepackage{amsmath, amssymb}

\begin{document}

\title{Thermodynamic Properties of Universal Fermi Gases}

\author{Erik M. Weiler and Theja N. De Silva}

\affiliation{Department of Physics, Applied Physics and Astronomy,
The State University of New York at Binghamton, Binghamton, New York
13902, USA.}

\begin{abstract}

We develop a simple, mean-field-like theory for the normal phase of a unitary Fermi gas by deriving a self-consistent equation for its self-energy via a momentum-dependent coupling constant for both attractive and repulsive universal fermions. For attractive universal fermions in the lower branch of a Feshbach resonance, we use zero-temperature Monte Carlo results as a starting point for one-step iteration in order to derive an analytical expression for the momentum-dependent self-energy. For repulsive universal fermions in the upper branch of a Feshbach resonance, we iteratively calculate the momentum-dependent self-energy via our self-consistent equation. Lastly, for the case of population imbalance, we propose an ansatz for higher order virial expansion coefficents. Overall, we find that our theory is in good agreement with currently available, high temperature experimental data.

\end{abstract}

\maketitle

\section{I. Introduction}

Fermion interactions play a central role in a wide range of physical systems. When these interactions are sufficiently strong, the physical properties of systems such as high-$T_c$ superconductors, neutron stars and cold atomic gases begin to exhibit universal behavior. Thanks to recent experimental progress, atoms and molecules can be trapped and cooled within optical or magnetic traps, which provide clean and highly-controllable environments where cold atomic systems can be studied. One such system, the two-component Fermi gas, is an excellent candidate for the observation of strongly interacting phenomena. We note that for cold atomic gases, the interaction between atoms is varied by manipulating their many-body bound states via the technique known as Feshbach resonance (FBR).

At characteristic densities and ultracold temperatures, only isotropic and short-range \emph{s}-wave scattering between particles can take place. This scattering can be characterized by a single parameter, the \emph{s}-wave scattering length $a$. Experimentally, the \emph{s}-wave scattering length can be tuned by using FBR~\cite{fbr1, fbr2, fbr3, fbr4, fbr5, fbr6, fbr7, fbr8}. In the unitarity limit, where $a$ is tuned to $\pm \infty$, the system is strongly interacting and its physical properties are independent of the shape of the inter-particle potential. As such, the system is expected to exhibit universal properties~\cite{ug1, ug2} since its corresponding equilibrium properties depend only on the scaled temperature $T/T_F$ (which is set by the energy scale $E_F$ and length scale $l$), where $T_F$, $E_F$, and $l$ are the Fermi temperature, the Fermi energy, and the inter-particle distance, respectively. Thus, by studying unitary Fermi gases, one learns about the equation of state for strongly interacting systems in general.

Thanks to the elegant, but simple, mean-field-like theories proposed by Eagle and Legget and Nozieres and Schmitt-Rink, zero-temperature superfluid properties are well understood~\cite{nsr1, nsr2, nsr3}. In the attractive interaction regime (where $a < 0$), atoms form Bardeen, Cooper, and Schreifer (BCS) pairs such that the ground state is a BCS superfluid. In the repulsive interaction regime (where $a > 0$), the atomic potential supports a two-body molecular bound state in vacuum such that the ground state is a Bose-Einstein Condensate (BEC) of these molecules. In between these two ground states, there is a smooth crossover where $a$ changes sign as it passes through $\infty$. Since the thermodynamics also evolve smoothly from the BCS limit to the BEC limit, a good understanding of the crossover regime comes from various interpolating schemes between these two limits~\cite{randeria}. However, due to the lack of theoretical techniques for taking into account strong interaction effects, the study of finite temperature, non-superfluid (or normal) phases are challenging. In particular, the quantitative theoretical understanding of the strongly correlated problem in general is limited by the absence of any small physical parameters within the unitarity limit. While several heavily numerical approaches~\cite{nmug1, nmug2, nmug3, nmug4, nmug5} have been used to resolve this issue, the simplicity of mean-field theory has been lost in the process.

In this paper, we develop a simple, mean-field-like theory for strongly interacting, normal-phased Fermi gases at the unitarity limit. To do so, we begin in Section II by constructing a self-consistent theory to determine the self-energy of a spin-balanced, unitary Fermi gas, which can be accomplished by calculating the total energy, entropy and variational occupation numbers of the system. Continuing in Section II, we explore the thermodynamics of both the "upper branch" and the "lower branch" of a FBR by calculating the finite temperature equation of state, such that the pressure and the entropy of the system can be extracted and compared with experimental data. Finally, in Section III, we generalize an accurate theory concerning population balanced fermions to the case of population imbalanced fermions at the unitarity limit. This is done by writing the virial coefficients for population imbalanced fermions in terms of the virial coefficients for population balanced fermions, thereby allowing us to determine the grand thermodynamic potential for a system of population imbalanced, unitary fermions.

\section{II. A Self-Consistent Theory for Universal Thermodynamics}

In general, interaction effects in Fermi gases can be well described by including the appropriate self-energy term~\cite{se1, se2} in an expression for the total energy of the particles. Despite the presence of strong interaction, it has been argued that such a self-energy exists for unitary Fermi gases~\cite{stf1}, however, it is impossible to derive an expression for it from the first principles of a Fermi gas alone. The main difficulty with constructing a self-energy for a degenerate, unitary Fermi gas is related to the strong interaction associated with an infinite \emph{s}-wave scattering length. To circumvent this problem, we start by writing the self-energy $\Sigma(k)$ in terms of a distribution function $n_k$ and a momentum-dependent coupling constant $g_{kk^\prime}$ as:

\begin{eqnarray}
\Sigma(k) = \frac{1}{2V}\sum_{k^{\prime}} g_{kk^\prime} n_{k^\prime}\label{SELFEN}
\end{eqnarray}

\noindent where $V$ is the system volume. Here, the distribution function $n_k$ is taken as a variational function. In this Hartree-Fock-like self-energy, $g_{kk^\prime}$ is given by:

\begin{eqnarray}
g_{kk^\prime} = -\frac{4\pi\hbar^2}{m} \frac{\delta(|k-k^\prime|/2)}{ |k-k^\prime|/2}\label{MDCC}
\end{eqnarray}

\noindent where the phase shift $\delta$ and the scattering length $a$ are related to one another by: $\delta = -\arctan(|k-k^\prime|a/2)$. Here, $m$ is the mass of a particle and $|k-k^\prime|/2 \equiv q$ is the relative momentum between a particle and the scattering particle. Note that in the limit where $(k-k^\prime)|a|/2 \ll 1$, the momentum-dependent coupling constant transforms into the ordinary momentum-independent, mean-field coupling contasnt $g = 4\pi\hbar^2 a/m$, while at the unitarity limit (where $a \rightarrow \pm \infty$), $\delta \rightarrow \delta_{0} = \mp \pi/2$, a \emph{constant} value.

Now, by calculating the total energy:

\begin{eqnarray}
E = \sum_k\frac{\hbar^2k^2}{2m} n_k + \frac{1}{2V}\sum_{k, k^\prime} g_{k, k^\prime} n_k n_{k^\prime}
\end{eqnarray}

\noindent the entropy:

\begin{eqnarray}
S = -\sum_k[n_k \ln n_k + (1-n_k) \ln (1-n_k)]
\end{eqnarray}

\noindent and the number of particles: $N = 2\sum_k n_k$, the grand thermodynamic potential $\Omega = E-TS-\mu N$ is derived. Then, by using the relation: $N = -\partial \Omega/\partial \mu$, the variational occupation number $n_k$ is obtained as:

\begin{eqnarray}
n_k = \frac{1}{e^{\beta(\epsilon_k-\mu)}+1}
\end{eqnarray}

\noindent where $\epsilon_k = \hbar^2k/2m+\Sigma(k)$ is the total single particle energy of a particle with momentum \emph{k} and $\mu$ is the chemical potential of the system. Lastly, by inserting this occupation number and the unitary value of the momentum-dependent coupling constant into Eq. (\ref{SELFEN}), we complete the derivation of a self-consistent equation for the self-energy $\Sigma(k)$ at unitarity:

\begin{eqnarray}
\Sigma(k) = -\frac{1}{2V}\sum_{k^{\prime}}\frac{8\pi\hbar^2}{m} \frac{\delta_{0}}{|k-k^\prime|}\frac{1}{e^{\beta(\epsilon_{k^\prime}-\mu)}+1}\label{SELFEN1}
\end{eqnarray}

However, since real-space densities for ultracold Fermi gases are so dilute, they also have large momentum-space densities associated with them. As a result, we can convert the sum in Eq. (\ref{SELFEN1}) into an equivalent integral over all of momentum-space. Then, by defining the dimensionless variables: $\beta \mu \equiv \eta$, $\beta \Sigma(k)\equiv V(\gamma)$ and $\gamma \equiv \hbar k\sqrt{\beta/(2m)}$ (where $\beta = 1/k_B T$), we have:

\begin{eqnarray}
V(\gamma) = -\frac{\delta_{0}}{\pi^2}\int \frac{{\gamma^{\prime}}^2 d{\gamma^{\prime}}}{e^{{\gamma^{\prime}}^2-\eta+V({\gamma^{\prime}})}+1} \frac{sin\theta^\prime d\theta^\prime d\phi^\prime}{|\gamma-{\gamma^{\prime}}| \label{SELFEN2}}
\end{eqnarray}

\noindent By expanding 1/$|\gamma - \gamma^\prime|$ in terms of spherical harmonics and using their orthonormal properties, we can solve the angular part of Eq. (\ref{SELFEN2}) for both the $\gamma < \gamma^{\prime}$ case and the $\gamma > \gamma^{\prime}$ case. Hence, we obtain:

\begin{eqnarray}
\int \frac{sin\theta^\prime d\theta^\prime d\phi^\prime}{|\gamma-\gamma^\prime|}=\begin{cases} 4\pi/\gamma^\prime, & \mbox{for } \gamma < \gamma^\prime \\ 4\pi/\gamma, & \mbox{for } \gamma > \gamma^\prime\end{cases}
\end{eqnarray}

\noindent Finally, we distribute Eq. (\ref{SELFEN2}) into two, separate integrals (corresponding to the two cases) and simplify the results to arrive at the concluding form of our self-consistent equation for the self-energy $V(\gamma)$ at unitarity:

\begin{eqnarray}
V(\gamma) = -\frac{4\delta_{0}}{\pi}\biggr[\frac{1}{\gamma}\int_0^\gamma\frac{y^2 dy}{e^{y^2-\eta+V(y)}+1} + \int_\gamma^\infty \frac{y dy}{e^{y^2-\eta+V(y)}+1}\biggr]\label{IESE}
\end{eqnarray}

\subsection{Upper Branch Thermodynamics of a Feshbach Resonance}

\begin{figure}
\includegraphics[width=\columnwidth]{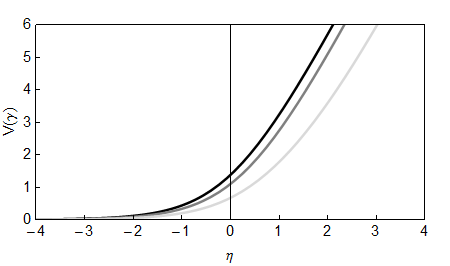}
\caption{The self-energy of a unitary Fermi gas in the upper branch of a Feshbach resonance as related to its momentum. Values of $\gamma = 0.1$, 1 and 2 are shown in black, dark gray and light gray lines, respectively.} \label{ubse}
\end{figure}

\begin{figure}
\includegraphics[width=\columnwidth]{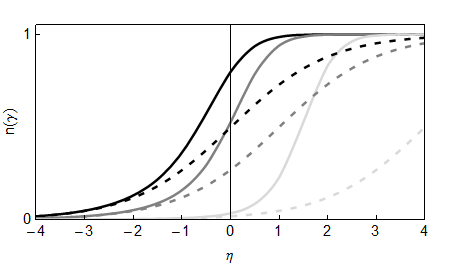}
\caption{The occupation numbers of a unitary Fermi gas in the upper branch of a Feshbach resonance as related to momentum. Values of $\gamma = 0.1$, 1 and 2 are shown in black, dark gray and light gray lines, respectively. (Dashed lines of the same hue are the \emph{non}-interacting occupation numbers corresponding to these $\gamma$ values.)} \label{occnum}
\end{figure}

\begin{figure}
\includegraphics[width=\columnwidth]{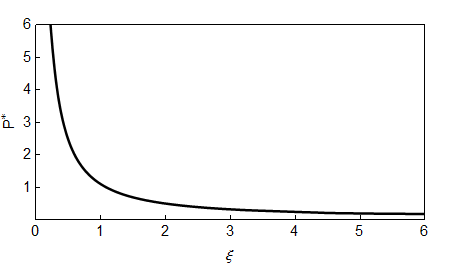}
\caption{The pressure of a unitary Fermi gas as a function of $\xi$ in the upper branch of a Feshbach resonance. Here, $P^*=P(\xi)\lambda^3/k_B T$, where $\xi = \exp(-\mu/k_BT)$.} \label{pressure}
\end{figure}

The BEC side of a Feshbach resonance is characterized by a positive scattering length ($a >0$) and by a ground state which involves bound pairs of fermionic molecules in condensate. This is known as the "lower branch" of a Feshbach resonance. By contrast, in the "upper branch" of a Feshbach resonance, the system's wavefunction consists of scattering states such that we can neglect these bound pairs and their corresponding binding energies~\cite{jump1}. Motivated by a series of recent experiments conducted in this (metastable) upper branch state \cite{jump2, exub1, exub2}, static and dynamic properties of Fermi gases have been theoretically studied \cite{thub1, thub2, thub3, thub4, thub5}. To further explore the thermodynamics of the upper branch, we set $\delta_{0}$ = $-\pi/2$ (i.e. its limiting value as $a \rightarrow +\infty$) in order to find the numerical solution of Eq. (\ref{IESE}) iteratively. We note that while Eq. (\ref{IESE}) converges very rapidly at first, it requires more iterations at relatively small values of $\gamma$. As such, the calculated self-energy $V(\gamma)$ for the upper branch is shown in FIG.~\ref{ubse} for three different values of $\gamma$, while the calculated occupation numbers $n(\gamma)$ for the upper branch are similarly shown in FIG.~\ref{occnum}. Additionally, we show the calculated pressure $P^*=P(\xi)\lambda^3/k_B T$ for the upper branch in FIG.~\ref{pressure}, where $\xi = \exp(-\mu/k_BT)$. It is important to note that the tail of the momentum distribution should decay with $\gamma^{-4}$ behavior as related to the Tan Contact~\cite{contact1, contact2, contact3}. However, the Contact density as a function of temperature was recently determined by Boettcher \emph{et al.} for the self-energy of ultracold fermions in the BEC-BCS crossover by using methods in non-perturbative quantum field theory~\cite{contact}. Their findings indicate that the Contact density is not a monotonic function of temperature, and that its maximum occurs at approximately 1.25$T_{c}$ (where $T_{c}$ is the critical temperature for phase transition). Additionally, their results show that the Contact density becomes very small for temperatures greater than 2$T_{c}$, which is reinforced by the findings of Enss and Haussmann, who have determined that the Contact density for a unitary Fermi gas is $C = 0.086k_{F}^4$ at $T = 0.5T_{F}$~\cite{contactd}. Since our theory is valid for the normal phase of a Fermi gas at unitarity, we believe that the Contact plays a very small role in the actual decay of our momentum distribution function, especially at higher temperatures.

\subsection{BCS Side Thermodynamics of a Feshbach Resonance}

The BCS side of a Feshbach resonance is characterized by a negative scattering length ($a < 0$) and by a ground state which involves Cooper pairs of fermionic molecules in condensate. To further explore the thermodyanmics of this lower branch state, we set $\delta_{0}$ = $+\pi/2$ (i.e. its limiting value as $a \rightarrow -\infty$) in order to find the numerical solution of Eq. (\ref{IESE}) iteratively. However, upon doing so, we realize that Eq. (\ref{IESE}) does \emph{not} converge for relatively small values of $\gamma$ as it did before (for small values of $\xi$). In principle, we could introduce a lower cutoff value for the momentum to get around this problem, but our resulting numerical iteration may not be reliable. As such, we choose to solve Eq. (\ref{IESE}) via one-step iteration instead. The motivation for doing so is two-fold: not only do we want to get away with the non-convergency at low values of momenta, but we also want to find an approximate analytical expression for the self-energy. However, obtaining an accurate expresssion for the self-energy requires us to first choose an accurate starting point for our one-step iteration (i.e. the zeroth order self-energy).

In general, the unitarity Fermi gas at zero-temperature has been studied with both the heavily numerical Monte Carlo method~\cite{nmug4} and with renormalization group theory~\cite{stf2}. Monte Carlo calculations at zero-temperature have shown that the self-energy is given by: $\hbar \Sigma(k) = A \mu$ (where $A \simeq -0.4045$~\cite{nmug2, lobo, chevy}), while renormalization group theory has shown that the self-energy has a weak momentum and frequency dependence at the unitarity limit~\cite{stf2}. Based on these zero-temperature theoretical results, we make the ansatz $\hbar \Sigma(k) = A \mu$ as the zeroth order self energy. Thus, for one-step iteration, we define: $h = \eta - V \equiv \eta - A\eta$. Doing so (while expanding the denominator) allows us to write the first integral on the right hand side of Eq. (\ref{IESE}) as:

\begin{eqnarray}
I_1 = \sum_{n=0}^\infty (-1)^n \int_0^{\sqrt{h}} e^{-n(h-y^2)} y^2 dy \\ \nonumber
+ \sum_{n=1}^\infty (-1)^{n+1} \int_{\sqrt{h}}^{\gamma} e^{-n(y^2-h)} y^2 dy \label{I1a}
\end{eqnarray}

\noindent Completing these two integrations yeilds the expression:

\begin{eqnarray}
I_1 = \sum_{n=0}^\infty (-1)^n g_n(\sqrt{h}) + \sum_{n=1}^\infty (-1)^{n+1}[f_n(\gamma) - f_n(\sqrt{h})] \label{I1b}
\end{eqnarray}

\noindent where $g_n(x) = e^{-nh}[2e^{nx^2} x \sqrt{n} - \sqrt{\pi} Erfi(x\sqrt{n}]/(4 n^{3/2})$ and $f_n(x) = e^{nh}[-2e^{-nx^2} x \sqrt{n} + \sqrt{\pi} Erf(x\sqrt{n}]/(4 n^{3/2})$. The two functions: $Erf(x)$ and $Erfi(x)$, are the usual error and imaginary error functions, respectively. Note that for $h <0$, only the second term in Eq. (\ref{I1b}) contributes, whereas for $h > \gamma$, only the first term contributes. Thus, for $h <0$, we must set $f_n(h\rightarrow 0) = 0$ in the second term. The second integral on the right hand side of Eq. (\ref{IESE}) can be evaluated explicitly to obtain:

\begin{eqnarray}
I_2 = (\ln[e^h + e^{\gamma^2}] -\gamma^2)/2
\end{eqnarray}

\begin{figure}
\includegraphics[width=\columnwidth]{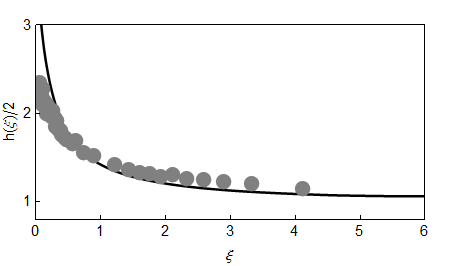}
\caption{The pressure of a unitary Fermi gas as a function of $\xi$ in the lower branch of a Feshbach resonance. Here, $h(\xi) = P(\mu,T)/ P_1(\mu,T) $, where $\xi = \exp(-\mu/k_BT)$. Our theory (black line) vs. experimental data~\cite{ens} (gray points).} \label{hf}
\end{figure}

Finally, within one-step iteration, we find an analytical expression for the momentum-dependent, finite temperature self-energy as: $V = -2(I_1/\gamma + I_2)$. Note that while we have completed the momentum integration, the self-energy now has the form of an infinite converging series. Now, in order to test the accuracy of our iterative analytical theory, we calculate the finite temperature equation of state and then compare our results with experimental data. In the unitary limit, the pressure can be written in a universal form as: $P(\mu,T) = P_1(\mu,T)h(\xi)$. Here, $P_1(\mu,T)$ is the pressure of a single-component, non-interacting Fermi gas:

\begin{eqnarray}
P_1(\mu,T) = \frac{k_B T}{\lambda^3}\frac{2}{\sqrt{\pi}}\int_0^\infty \sqrt{t}\ln[1+z_\sigma e^{-t}] dt
\end{eqnarray}

\noindent where $\lambda=\sqrt{2\pi \hbar^2/m k_B T}$ is the thermal wavelength and $\xi = \exp(-\mu/k_BT)$. Upon completion of the self-energy calculation, we compute the system pressure via the relation: $P(\mu,T) = -\partial \Omega/\partial V$, and then extract the universal function $h(\xi)$. This function is plotted in FIG.~\ref{hf} together with experimental data from Nascimbene \emph{et al.}~\cite{ens}. As can be seen, our theory is in reasonable agreement with experimental values at higher temperatures, however, we begin to see a noticeable deviation at very low temperatures. This is due to the fact that our theory is only valid for the \emph{normal} phase, wheras the experimental system is in the superfluid phase.

We note that while the jump in $\delta_{0}$ by $\pi$ at unitarity will correspond to a jump in our self-energy, the thermodynamic quantities of the system will remain \emph{continuous} throughout the BEC-BCS evolution. This is due to our treatment of the upper branch on the BEC side, where we have chosen to neglect the binding energy of ground-state pairs. As such, the jump in our self-energy should vanish if we were to take these binding energies into account. However, since this was not the case, FIG.~\ref{pressure} and FIG.~\ref{hf} are qualitatively the same, but quantitatively \emph{different}. Note that this behavior has already been experimentally verified via upper branch energy measurements~\cite{jump2}, thus we believe that these two figures would also be quantitatively the same if the jump in our self-energy was not present. Regardless, we have chosen to validate our theory by using data~\cite{ens} from experiments performed on the BCS side of a Feshbach resonance, where its numerical predictions are most suited for comparison with thermodynamic measurements.

Similarly, the temperature dependence of the entropy and energy for harmonically trapped, unitary fermions is measured by Thomas's group at Duke University~\cite{duke}. To investigate this, we used the local density approximation (LDA) to evaluate the entropy and energy of these trapped fermions. In LDA, the local chemical potential $\mu$ is written in terms of the central chemical potential $\mu_0$ and the trapping potential $V(\vec{r}) = m[\omega_\bot^2(x^2+y^2)+\omega_z^2z^2]/2$ as: $\mu = \mu_0 -m\omega^2 r^2/2$, where $\omega = (\omega_\bot^2\omega_z)^{1/3}$ is the average trapping frequency. In doing so, the total number of particles:

\begin{eqnarray}
N = \int d^3\vec{r} n(\vec{r}) = \frac{4\pi}{m\omega^2}\int P(r) dr
\end{eqnarray}

\noindent and the total energy:

\begin{eqnarray}
E &=& 12 \pi\int r^2 P(r) dr
\end{eqnarray}

\noindent can easily be converted into integrals over the chemical potential as:

\begin{eqnarray}
N &=& \frac{4\pi}{\sqrt{2\beta m^3\omega^6}}\int  \frac{P(\eta) d\eta}{\sqrt{\eta_0-\eta}}\label{ne} \\
E& = & \frac{12\sqrt{2}\pi}{\sqrt{\beta^3 m^3\omega^6}}\int P(\eta) \sqrt{\eta_0-\eta}d\eta \label{ee},
\end{eqnarray}

\noindent where $\eta = \beta \mu$ and $\eta_0 = \beta \mu_0$. The entropy is then given by: $S = 4E\beta/3 - \eta_0N$. Now, by defining the Fermi energy and Fermi temperature of an ideal Fermi gas as: $E_F \equiv k_BT_F = (3N)^{1/3}\hbar \omega$, we can combine Eq. (\ref{ne}) and (\ref{ee}) to yeild the expression:

\begin{eqnarray}
\frac{S}{Nk_B} &=& \frac{4}{3}\frac{T_F}{T}\frac{E}{NE_F}.
\end{eqnarray}

\begin{figure}
\includegraphics[width=\columnwidth]{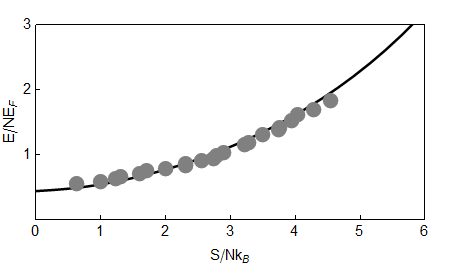}
\caption{The entropy as a function of energy for harmonically trapped, unitary fermions. Our theory (black line) vs. experimental data~\cite{duke} (gray points).} \label{se}
\end{figure}

\noindent Hence, for given values of $\eta_0$ and $T$, we can solve the above set of equations for the entropy and the energy. As such, the calculated entropy as a function of energy is shown in FIG. \ref{se} along with experimental data taken from Luo and Thomas~\cite{duke}.

\section{III. Population Imbalanced Fermions at Unitarity}

Recent experiments concerning population imbalanced fermions~\cite{ru1, mit1} has triggered a new direction in theoretical research devoted to the study of unitary fermions in the presence of population imbalance~\cite{stf1, fc, liu}. For population balanced, two-component fermions at unitarity, R. K. Bhaduri, W. van Dijk and M. V. N. Murthy (BvDM) have proposed a parameter-free, high-temperature equation of state based on a virial cluster expansion~\cite{BvDM} that shows excellent agreement with experimental results over a wide range of fugacity. Their basic assumption is that higher order cluster integrals can be written in terms of two-particle clusters. This is justified because only two-body scattering effects are dominant for dilute atomic gases, even at unitarity where the virial coefficients are temperature independent. In this section, we generalize the BvDM approach to the case of population imbalanced, unitary fermions.

First, we summarize the original BvDM approach~\cite{BvDM} by noting that the grand thermodynamic potential of a population balanced Fermi system can be written as:

\begin{eqnarray}
\Omega - \Omega^{(0)} = -k_BTZ_1(\beta) \sum_{l=0}^\infty (\Delta b_l)z^l,
\end{eqnarray}

\noindent where $\Omega^{(0)}$ is the grand thermodynamic potential of an ideal Fermi gas, $Z_1(\beta)$ is the one-particle partition function and $\Delta b_l = b_l-b^{(0)}_l$ is the $l$-particle cluster integral relative to an ideal Fermi gas. As FBR is related to the forming and dissolving of two-body pairs, BvDM has proposed that higher order cluster integrals are expressible in terms of the two-body cluster $\Delta b_2$. Assuming the $l$-body cluster is one particle interacting with $l-1$ paired particles, the $l$-particle cluster integral is given by:

\begin{eqnarray}
\Delta b_l = (-1)^l \frac{\Delta b_2}{2^{\alpha_l}}
\end{eqnarray}

\noindent for $l \ge 2$, where $\alpha_l = (l-1)(l-2)/2$. As was mentioned earlier, the BvDM ansatz for a population balanced system of fermions shows excellent agreement with experimental results over a wide range of fugacity. Hence, for the remainder of this paper, we generalize this ansatz to the case of population imbalanced, unitary fermions.

The grand thermodynamic potential of a population imbalanced Fermi system can be written as:

\begin{eqnarray}
\Omega = -k_BTZ_1(\beta) \sum_{n=1}^\infty \sum_{k=0}^\infty b_{n,k}z_\uparrow^{n-k}z_\downarrow^k
\end{eqnarray}

\noindent where $b_{n,k}$ is the $n$-th virial coefficient for a configuration of $n-k$ spin-up fermions and $k$ spin-down fermions. We also note that $b_{n,k}$ has the properties: $b_{n,n-k} = b_{n,k}$ and $\sum_{k=0}^n b_{n,k} = b_n$. Now, by defining the virial coefficient difference relative to non-interacting fermions as: $\Delta b_{n,k} = b_{n,k}-b_{n,k}^{(0)}$, we have:

\begin{eqnarray}
\Omega - \Omega_0 = -k_BTZ_1(\beta) \sum_{n=1}^\infty \sum_{k=0}^\infty (\Delta b_{n,k})z_\uparrow^{n-k}z_\downarrow^k,
\end{eqnarray}

\noindent where $\Omega_0 = \sum_\sigma \Omega_{0\sigma}$. Here, we note that $\Omega_{0\sigma}$ is the grand thermodynamic potential for $\sigma$ non-interacting fermions, and is given by the expression:

\begin{eqnarray}
\Omega_{0\sigma}= -V\frac{k_B T}{\lambda^3}\frac{2}{\sqrt{\pi}}\int_0^\infty \sqrt{t}\ln[1+z_\sigma e^{-t}] dt.
\end{eqnarray}

\noindent Since the interaction occurs only between fermions with opposing spins, the virial coefficients for population imbalanced fermions can be written in terms of virial coefficients for population balanced fermions. Furthermore, $\Delta b_{n,0} = 0$, and we find that: $\Delta_{n,k} = \Delta_n/(n-1)$ for $n \ge 2$. Finally, by putting everything together, the grand thermodynamic potential for a population imbalanced, unitary Fermi system can be written as:

\begin{eqnarray}
\Omega - \Omega_0 = -k_BTZ_1(\beta)\sum_{n=2}^\infty \frac{(-1)^n \Delta b_2}{2^{\alpha_n}(n-1)} \sum_{k=1}^{n-1}z_\uparrow^{n-k}z_\downarrow^k\label{omega}
\end{eqnarray}

\noindent Hence, by using the relation: $P = -\partial \Omega/\partial V$, we see that the pressure for a population imbalanced, unitary Fermi system can be similarly written as:

\begin{eqnarray}
P - P_0 = \frac{k_B T}{\lambda^3} \sum_{n=2}^\infty \frac{(-1)^n \Delta b_2}{2^{\alpha_n}(n-1)} \sum_{k=1}^{n-1}z_\uparrow^{n-k}z_\downarrow^k\label{pres}
\end{eqnarray}

\begin{figure}
\includegraphics[width=\columnwidth]{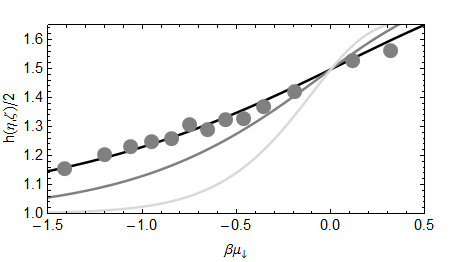}
\caption{The finite temperature equation of state for a population imbalanced, unitary Fermi gas. Values of $\eta = 1.0$, 0.5 and 0.25 are shown in black, dark gray and light gray lines, respectively. The gray points represent experimental data~\cite{ens} for the population balanced case, which is recovered from Eq. (\ref{hh}) when $\eta = 1.0$.} \label{pita}
\end{figure}

\noindent Lastly, we extract the universal function $h(\eta,\xi)$ from Eq. (\ref{pres}) to obtain:

\begin{eqnarray}
h(\eta,\xi) = 1 + \frac{1}{\Omega_0}\sum_{n=2}^\infty \frac{(-1)^n \Delta b_2}{2^{\alpha_n}(n-1)} \sum_{k=1}^{n-1}z_\uparrow^{n-k}z_\downarrow^k\label{hh}
\end{eqnarray}

\noindent where  $\eta\equiv\mu_\downarrow/\mu_\uparrow$ is the ratio between the two chemical potentials. We note that when $\eta = 1.0$, the chemical potentials are equal and Eq. (\ref{hh}) reduces to its population balanced form as depicted in FIG. \ref{hf}. By contrast, $h(\eta,\xi)$ is plotted in FIG. \ref{pita} as a function of $\beta\mu_\downarrow$ for three different values of $\eta$, along with experimental data~\cite{ens} pertaining to the population balanced case for verification purposes.

\section{IV. Conclusions and Remarks}

Two important points of interest to consider are the recent experimental findings by Zwierleins's group~\cite{imb1} and the recent theoretical findings by Van Houcke \emph{et al}~\cite{imb2}. Zwierlein's group observed the superfluid phase transition in a strongly-interacting Fermi gas by using high-precision measurements of the local compressibility, density and pressure. Their data was able to completely describe the universal thermodyanmics of such Fermi gases without the use of any fit or external thermometer. Similarly, Van Houcke \emph{et al.} computed and measured the equation of state for a normal, unitary Fermi gas. Their data showed excellent agreement with their theory that a series of Feynman diagrams can be controllably resummed in a non-perturbative regime using a Bold Diagrammic Monte Carlo approach. We note that while the newer data from Zwierlein's group is highly accurate~\cite{imb1}, its normal phase measurements are nearly identical to the measurements made by Nascimbene \emph{et al.}~\cite{ens}.

In conclusion, we have presented a self-consistent theory to determine the self-energy of a strongly interacting, normal-phased Fermi gas at unitarity. We have also shown that this self-energy can be used to calculate universal thermodynamic properties of a Fermi gas for both the upper branch of a Feshbach resonance and the lower branch of a Feshbach resonance. In addition, we have demonstrated that higher order virial expansion coefficients for population imbalanced fermions can be written in terms of virial expansion coefficients for population balanced fermions, which makes calculating the grand thermodynamic potential for population imbalanced Fermi systems at unitarity a much less cumbersome task. Overall, we find that our theory is in good agreement with currently available experimental data, indicating that there may be promising advancements ahead for further theoretical research regarding fermion interactions at the unitarity limit.

\section{V. Acknowledgements}

We are grateful to Sylvain Nascimbene for sending us experimental data regarding their universal function measurements, and we would like to thank John Thomas for pointing out their experimental data in Ref.~\cite{duke}.


\begin{references}

\bibitem{fbr1} M. Houbiers, H. T. C. Stoof, W. I. McAlexander, and R. G. Hulet, Phys. Rev. A \textbf{57}, R1497 (1998).

\bibitem{fbr2} K. M. O'Hara, S. L. Hemmer, M. E. Gehm, S. R. Granade and J. E. Thomas, Science \textbf{298}, 2179 (2002).

\bibitem{fbr3} K. Dieckmann, C. A. Stan, S. Gupta, Z. Hadzibabic, C. H. Schunck, and W. Ketterle, Phys. Rev. Lett. \textbf{89}, 203201 (2002).

\bibitem{fbr4} S. Jochim, M. Bartenstein, A. Altmeyer, G. Hendl, C. Chin, J. Hecker Denschlag, and R. Grimm, Phys. Rev. Lett. \textbf{91}, 240402 (2003).

\bibitem{fbr5} T. Bourdel, J. Cubizolles, L. Khaykovich, K. M. F. Magalhaes, S. J. J. M. F. Kokkelmans, G. V. Shlyapnikov, and C. Salomon, Phys. Rev. Lett. \textbf{91}, 020402 (2003).

\bibitem{fbr6} C. A. Regal and D. S. Jin, Phys. Rev. Lett. \textbf{90}, 230404 (2003).

\bibitem{fbr7} J. E. Thomas, J. Kinast, and A. Turlapov, PRL \textbf{95}, 120402 (2005).

\bibitem{fbr8} M. Horikoshi, S. Nakajima, M. Ueda, and T. Mukaiyama, Science \textbf{327}, 442 (2010).

\bibitem{ug1} H. Heiselberg, Phys. Rev. A, \textbf{63}, 043606 (2001).

\bibitem{ug2} S. Giorgini, L. P. Pitaevskii, and S. Stringari, Rev. Mod. Phys. \textbf{80}, 1215, (2008).

\bibitem{nsr1} D. M. Eagles, Phys. Rev. \textbf{186}, 456 (1969).

\bibitem{nsr2} A. J. Leggett, in Modern Trends in the Theory of Condensed Matter, edited by A. Pekalski and R. Przystawa (Springer-Verlag, Berlin, 1980).

\bibitem{nsr3} P. Nozieres and S. Schmitt-Rink, J. Low Temp. Phys. \textbf{59}, 195 (1985).

\bibitem{randeria} C. A. R. Sa de Melo, Mohit Randeria, and Jan R. Engelbrecht, Phys. Rev. Lett. \textbf{71}, 3202 (1993).

\bibitem{nmug1} R. Haussmann, M. Punk, and W. Zwerger, Phys. Rev. A \textbf{80}, 063612 (2009).

\bibitem{nmug2} C. Lobo, A. Recati, S. Giorgini, and S. Stringari, Phys. Rev. Lett. \textbf{97}, 200403 (2006).

\bibitem{nmug3} Piotr Magierski, Gabriel Wlazlowski, Aurel Bulgac, and Joaquin E. Drut, Phys. Rev. Lett. \textbf{103}, 210403 (2009).

\bibitem{nmug4} Evgeni Burovski, Nikolay Prokofev, Boris Svistunov, and Matthias Troyer, Phys. Rev. Lett. \textbf{96}, 160402 (2006).

\bibitem{nmug5} Joaquin E. Drut, Timo A. Lhde, Gabriel Wlazlowski, and Piotr Magierski, Phys. Rev. A \textbf{85}, 051601 (R) (2012).

\bibitem{se1} J. J. Kinnunen, Phys. Rev. A \textbf{85}, 012701 (2012).

\bibitem{se2} Pye-Ton How, Andre LeClair, J. Stat. Mech. P07001 (2010).

\bibitem{stf1} J. M. Diederix, H. T. C. Stoof, The BCS-BEC Crossover and the Unitary Fermi Gas, edited by W. Zwerger (Lecture Notes in Physics, Vol. 836, Springer, 2011).

\bibitem{jump1} Vijay B. Shenoy and Tin-Lun Ho, Phys. Rev. Lett. \textbf{107}, 210401 (2011).

\bibitem{jump2} T. Bourdel, J. Cubizolles, L. Khaykovich, K. M. F. Magalhaes, S. J. J. M. F. Kokkelmans, G. V. Shlyapnikov and C. Salomon, Phys. Rev. Lett. \textbf{91}, 020402 (2003).

\bibitem{exub1} G.-B. Jo, Y.-R. Lee, J.-H. Choi, C. A. Christensen, T. H. Kim, J. H. Thywissen, D. E. Pritchard, and W.
Ketterle, Science \textbf{325}, 1521 (2009).

\bibitem{exub2} Ye-Ryoung Lee, Myoung-Sun Heo, Jae-Hoon Choi, Tout T. Wang, Caleb A. Christensen, Timur M. Rvachov, Wolfgang Ketterle, e-print arXiv:1204.4229 (2012).

\bibitem{thub1} Vijay B. Shenoy, and Tin-Lun Ho, Phys. Rev. Lett. \textbf{107}, 210401 (2011).

\bibitem{thub2} S. Pilati, G. Bertaina, S. Giorgini, and M. Troyer, Phys. Rev. Lett. \textbf{105}, 030405 (2010).

\bibitem{thub3} S.-Y. Chang, M. Randeria, and N. Trivedi, Proc. Nat. Acad. Sci., \textbf{108}, 51 (2011).

\bibitem{thub4} L. J. Le Blanc, J. H. Thywissen, A. A. Burkov, and A. Paramekanti, Phys. Rev. A \textbf{80}, 013607 (2009).

\bibitem{thub5} Edward Taylor, Shizhong Zhang, William Schneider, Mohit Randeria, Phys. Rev. A \textbf{84}, 063622 (2011).

\bibitem{contact1} Shina Tan, Ann. Phys. \textbf{323}, 2952 (2008).

\bibitem{contact2} Shina Tan, Ann. Phys. \textbf{323}, 2971 (2008).

\bibitem{contact3} Shina Tan, Ann. Phys. \textbf{323}, 2987 (2008).

\bibitem{contact} Igor Boettcher, Sebastian Diehl, Jan M. Pawlowski and Christof Wetterich, e-print arXiv:1209.5641.

\bibitem{contactd} Tilman Enss and Rudolf Haussmann, e-print arXiv:1207.3103 (2012).

\bibitem{stf2} K. B. Gubbels and H. T. C. Stoof, Phys. Rev. Lett. \textbf{100}, 140407 (2008).

\bibitem{lobo} A. Recati, C. Lobo, and F. Chevy, Phys. Rev. Lett. \textbf{98}, 180402 (2007).

\bibitem{chevy} F. Chevy, Phys. Rev. A \textbf{74}, 063628 (2006).

\bibitem{ens} S. Nascimbene, N. Navon, K. J. Jiang, F. Chevy, and C. Salomon, Nature \textbf{463}, 1057 (2010).

\bibitem{duke} Le Luo and J. E. Thomas, J. Low Temp. Phys. \textbf{154}, 1 (2009).

\bibitem{ru1} G.B. Partridge, W. Li, R.I. Kamar, Y. Liao, and R.G. Hulet, Science, \textbf{311}, 503 (2006).

\bibitem{mit1} M.W. Zwierlein, A. Schirotzek, C.H. Schunck, and W. Ketterle, Science, \textbf{311}, 492 (2006).

\bibitem{fc} Frederic Chevy and Christophe Mora, Reports on Progress in Physics \textbf{73}, 112401 (2010).

\bibitem{liu} Xia-Ji Liu and Hui Hu, Phys. Rev. A \textbf{82} 043626 (2010).

\bibitem{BvDM} R. K. Bhaduri, W. van Dijk and M. V. N. Murthy, Phys. Rev. Lett. \textbf{108}, 260402 (2012).

\bibitem{imb1} Mark J. H. Ku, Ariel T. Sommer, Lawrence W. Cheuk, Martin W. Zwierlein, Science \textbf{335}, 563 (2012).

\bibitem{imb2} K. Van Houcke, F. Werner, E. Kozik, N. Prokof'ev, B. Svistunov, M. J. H. Ku, A. T. Sommer, L. W. Cheuk, A. Schirotzek and M. W. Zwierlein, Nature Phys. \textbf{8}, 366 (2012).

\end{references}
\end{document}